\documentstyle[aps,prl,epsf,floats]{revtex}

\draft

\begin{document}
\twocolumn[\hsize\textwidth\columnwidth\hsize\csname
@twocolumnfalse\endcsname

\title{New constraints on neutrino physics from Boomerang data}

\author{Steen Hannestad}

\address{NORDITA, Blegdamsvej 17, DK-2100 Copenhagen, Denmark}

\date{May 1, 2000}

\maketitle

\begin{abstract}
We have performed a likelihood analysis of the recent data on the
Cosmic Microwave Background Radiation (CMBR) anisotropy taken by the
Boomerang experiment.  We find that this data places a strong upper
bound on the radiation density present at recombination. Expressed in
terms of the equivalent number of neutrino species the $2\sigma$ bound
is $N_\nu \leq 13$, and the standard model prediction, $N_\nu = 3.04$,
is completely consistent the the data.  This bound is complementary to
the one found from Big Bang nucleosynthesis considerations in that it
applies to any type of radiation, i.e.  it is not flavour
sensitive. It also applies to the universe at a much later epoch, and
as such places severe limits on scenarios with decaying neutrinos. The
bound also yields a firm upper limit on the lepton asymmetry in the
universe.
\end{abstract}

 \vskip1.9pc]


{\it Introduction}--- The standard Big Bang model has been remarkably
successful in describing the observed features in our universe
\cite{peacock}.  The latest and most impressive confirmation of the
model comes from observations of anisotropies in the cosmic microwave
background radiation (CMBR). These anisotropies are predicted by the
Big Bang model, and were first discovered by the COBE satellite in
1992 \cite{COBE}.  Subsequently it was realized that precision
measurements of these anisotropies can yield very detailed information
about the fundamental cosmological parameters \cite{boef}, and
accordingly a vast number of papers have investigated the potential of
upcoming experiments to measure these parameters (see for instance
\cite{jungman1,jungman,tegmark1,bond1,eisenstein} and references
therein).

Now we have the first results which may rightly be called precision
CMBR measurements. They stem from the balloon-borne experiment
Boomerang which was flown over the Antarctica in 1999
\cite{bernardis}.

The results indicate that the universe is flat, and are essentially a
confirmation of the standard Big Bang model
\cite{bernardis,white2000,tegmark2}.

The data is so good that it can also be used to constrain physics
beyond the particle physics standard model.  A strong indication of
such exotic physics would be additional radiation energy in the
universe at the time of recombination.  This could for instance be
caused by additional light neutrinos, or some other exotic particle
which decoupled at very high temperature \cite{peacock}.  In the
present letter we use the data from Boomerang to place a strict upper
limit on the radiation density present at the time of recombination
($T \simeq 1$ eV).  The standard way of expressing the energy density
in light, non-interacting species, is in terms of the equivalent
number of neutrinos \cite{peacock}
\begin{equation}
N_{\rm eff, i} = \frac{\rho_i}{\rho_{\nu,0}},
\end{equation}
where $\rho_{\nu,0}$ is the energy density in a standard neutrino
species.

From Big Bang nucleosynthesis one can also infer a limit to the
effective number of neutrino species. By observing the primordial
abundances of D, $^4$He and $^7$Li, and comparing them to the
theoretically predicted values, one can infer an upper limit to
$N_\nu$ of \cite{lisi}
\begin{equation}
N_{\nu, {\rm BBN}} \leq 4.
\end{equation}

The CMBR limit can be viewed as complementary to the BBN limit because
the limit from BBN applies to the radiation energy density present
when the temperature of the universe was about 1 MeV, whereas the CMBR
limit applies at a temperature of 1 eV.  Furthermore, the BBN limit is
also flavour sensitive.  If the extra energy density is in the form of
electron neutrinos, it changes the weak reaction rates for the beta
processes that interconvert neutrons and protons. With some fine
tuning, even a very substantial amount of energy can be hidden in the
neutrino sector while still yielding the same outcome for BBN.

That is not the case for CMBR. In this case, extra energy density is
detectable because the CMBR spectrum changes with the addition of
radiation.  After recombination, the CMBR fluctuations can still
change. If the universe is completely matter dominated and flat, the
photons see a constant gravitational potential (in the linear
approximation), and thus travel with constant energy, except for the
overall redshifting.  However, immediately after recombination the
universe was not completely matter dominated so that the gravitational
potential was not constant. This leads to an enhancement of the first
acoustic peak in the power spectrum and is known as the early
Integrated Sachs-Wolfe (ISW) effect \cite{tegmark1}.  This effect is
only sensitive to energy density and not to the specific nature of the
radiation.

{\it Likelihood analysis} --- The CMBR fluctuations are usually
described in terms of the power spectrum, which is again expressed in
terms of $C_l$ coefficients as $l(l+1)C_l$, where
\begin{equation}
C_l \equiv \langle |a_{lm}|^2\rangle.
\end{equation}
The $a_{lm}$ coefficients are given in terms of the actual temperature
fluctuations as
\begin{equation}
T(\theta,\phi) = \sum_{lm} a_{lm} Y_{lm} (\theta,\phi).
\end{equation}
Given a set of experimental measurements, the likelihood function is
\begin{equation}
{\cal L}(\Theta) \propto \exp \left( -\frac{1}{2} x^\dagger
[C(\Theta)^{-1}] x \right),
\end{equation}
where $\Theta = (\Omega, \Omega_b, H_0, n, \tau, \ldots)$ is a vector
describing the given point in parameter space. $x$ is a vector
containing all the data points and $C(\Theta)$ is the data covariance
matrix.  This applies when the errors are Gaussian. If we also assume
that the errors are uncorrelated, this can be reduced to the simple
expression, ${\cal L} \propto e^{-\chi^2/2}$, where
\begin{equation}
\chi^2 = \sum_{i=1}^{N_{\rm max}} \frac{(C_{l, {\rm obs}}-C_{l,{\rm
theory}})_i^2} {\sigma(C_l)_i^2},
\label{eq:chi2}
\end{equation} 
is a $\chi^2$-statistics and $N_{\rm max}$ is the number of power
spectrum data points \cite{oh}.  In the present letter we use
Eq.~(\ref{eq:chi2}) for calculating $\chi^2$.

The procedure is then to calculate the likelihood function over the
space of cosmological parameters. The likelihood function for $N_\nu$
is then obtained by keeping $N_\nu$ fixed and maximizing ${\cal L}$
over the remaining parameter space.  The fundamental free parameters
which we allow to vary are: $\Omega_m$, the matter density,
$\Omega_b$, the baryon density, $h$, the Hubble parameter, $n$, the
spectral index, $\tau$, the optical depth to reionization, and $Q$,
the overall normalization of the spectrum given in terms of the
quadrupole moment
\footnote{Note that there is an estimated 10\% calibration uncertainty
in the overall normalization of the Boomerang data. However, this
effect is completely degenerate with varying $Q$ because we use only
one data set. Therefore we do not need to worry about it.}.  The range
in which they are allowed to vary are given in Table I.  We assume a
flat universe with $\Omega_0 = \Omega_b + \Omega_m + \Omega_\Lambda =
1$. This is the value strongly suggested by the Boomerang experiment
\cite{bernardis}.  Also, the final result of the analysis does not
vary much even if $\Omega_0$ is allowed to vary. For a given value of
$N_\nu$ we maximize the likelihood over the remaining parameter space
by using the non-linear optimization method called simulated annealing
\cite{H1}.  The data set which we use is the one given in de Bernardis
{\it et al.} \cite{bernardis}, and we use the publicly available
CMBFAST package for calculating theoretical power spectra \cite{SZ96}.

\narrowtext
\begin{table}
\caption{The free parameters used in the present analysis, as well as
the range in which they are allowed to vary.}
\begin{tabular}{lc}
Parameter & range \\ \tableline $Q$ & 5-40 $\mu$K \\ $\Omega_m$ &
0.2-1 \\ $\Omega_b h^2$ & 0.002-0.050 \\ $h$ & 0.30-0.9 \\ $n$ &
0.7-1.3 \\ $\tau$ & 0-1
\end{tabular}
\end{table}

We show the result of the likelihood analysis in Fig.~1 in terms of
$\chi^2$.  The absolute minimum $\chi^2$ is 5.27 at $N_\nu = 1.6$.
The number of Boomerang data points is 12 \cite{bernardis}, and we
allow 6 parameters to vary.  Thus, the number of degrees of freedom in
our fit is $6$, so that the obtained best fit $\chi^2$ falls within
$1\sigma$ of the expected for an acceptable fit.  The CMBR constraint
on $N_\nu$ is given by
\begin{equation}
N_\nu \leq \cases{6.2 & $1\sigma$, \cr 13 & $2\sigma$.}
\end{equation}
While this constraint is clearly much weaker than the $N_\nu \leq 4$
obtained from BBN considerations \cite{lisi}, it applies to any type
of relativistic energy density.

\begin{figure}[h]
\begin{center}
\epsfysize=7truecm\epsfbox{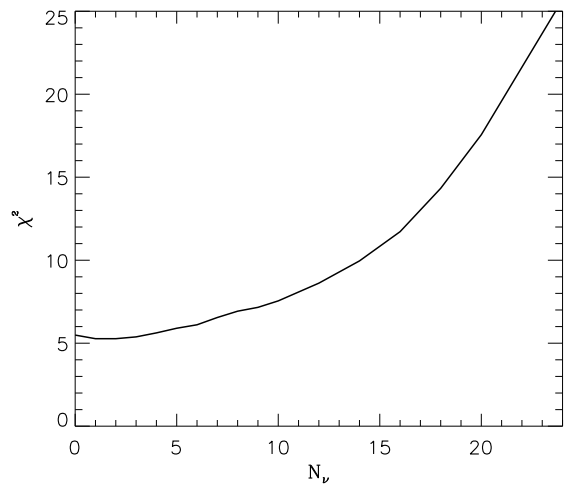}
\vspace{0truecm}
\end{center}
\vspace*{1cm}
\caption{$\chi^2$ for the Boomerang data as a function of $N_\nu$.
The curve has been obtained by minimizing $\chi^2$ over all other free
parameters.}
\label{fig1}
\end{figure}

{\it Discussion}--- We have calculated a strong upper bound on
radiation density present at recombination.  Next, we may ask what
bounds can be placed on neutrino properties from this.

The standard model prediction is that $N_\nu = 3.04$ \cite{lopez}.
Although the absolute minimum for $\chi^2$ is at $N_\nu=1.6$, the
standard model is completely consistent with the Boomerang data.
There is no indication in the data of neutrino physics beyond the
standard model.

If sterile neutrino degrees of freedom exist, then they can be exited
by oscillations in the early universe \cite{kainulai,H4}.  However, a
sterile neutrino can at most contribute $N_{\nu, {\rm sterile}}=1$ so
that our CMBR bound does not even exclude this possibility at the
$1\sigma$ level.

Next, we have no direct measurement of the lepton asymmetry in the
universe, and quite a large lepton asymmetry could in fact be hidden
in the neutrino sector \cite{freese,kang,lesgourgues,kinney}.  The
lepton asymmetry in neutrinos is usually expressed in units of
$\xi_\nu \equiv \mu_\nu/T_\nu$. The neutrino distribution function is
then given by $f = (\exp(E/T_\nu \pm \xi_\nu)+1)^{-1}$, where $+$
applies to neutrinos and $-$ to antineutrinos.
From BBN, one obtains the bound \cite{kang}
\begin{eqnarray}
\xi_{\nu_e} & \in & [-0.06,1.1] \\
|\xi_{\nu_{\mu,\tau}}| & \leq & 6.9.
\end{eqnarray}
This is in the absence of any oscillations.
For massless neutrinos, a chemical potential is equivalent to a change
in the effective number of species \cite{lesgourgues}
\begin{equation}
N_{\nu,{\rm eff}} = 3 + \frac{30}{7} \left(\frac{\xi}{\pi}\right)^2
+ \frac{15}{7} \left(\frac{\xi}{\pi}\right)^4.
\end{equation}
Using this, we can translate our $2\sigma$ upper bound on $N_\nu$ to 
a bound on $\xi_\nu$
\begin{equation}
|\xi_{\nu_{e,\mu,\tau}}| \leq 3.7 \,\,\, (2\sigma).
\end{equation}
This bound applies if only one species has a chemical potential. If
the asymmetry is equally shared then $|\xi_\nu| \leq 2.4$.
Lesgourgues and Peloso \cite{LP2000} have also discussed a cosmological
lepton asymmetry as a possible explanation of the relatively low
amplitude of the second acoustic peak in the Boomerang data, but they
did not perform a likelihood analysis of the data.

Interestingly this bound is at the brink of excluding a scenario for
the production of ultra high energy cosmic rays (UHECRs), proposed by
Gelmini and Kusenko \cite{gelmini}.  In this scenario, very high
energy neutrinos interact with a degenerate sea of neutrinos within
about 50 Mpc of earth. These interactions produce charged particles
such as protons which would then be the observed UHECR
primaries. However, because of the small $\nu \nu$ interaction cross
section this scenario only works if the cosmic neutrino background is
degenerate with $\xi \simeq 4$.  At present, the CMBR bound disfavours
this model, but cannot exclude it completely.  It has been estimated
that with the upcoming MAP and Planck experiments, it will be possible
to constrain $\xi$ with a precision of about 0.1 \cite{kinney}.  This
will definitively confirm or rule out scenarios like this.

Neutrino decays to massless secondaries prior to recombination are
also seen in the CMBR fluctuations as an increased $N_\nu$
\cite{H2,H3,lopez2,lopez3,lopez4}.  The effective number of neutrinos
is given roughly by
\begin{equation}
N_\nu \simeq 3 + 0.516 \alpha^{2/3},
\end{equation}
where $\alpha$ is the decay ``relativity'' parameter
\begin{equation}
\alpha \equiv 3.50 \left(\frac{m_\nu}{1 {\rm keV}}\right)^2
\left(\frac{\tau}{1 {\rm y}}\right).
\end{equation}
Again we can translate our bound on $N_\nu$ into a bound on $\alpha$
and thereby on neutrino decays
\begin{equation}
\alpha \leq 85.3 \,\,\, (2\sigma).
\label{eq:alpha}
\end{equation}
In Fig.~2 we show what this translates into in terms of neutrino
lifetime and mass.
\begin{figure}[h]
\begin{center}
\epsfysize=7truecm\epsfbox{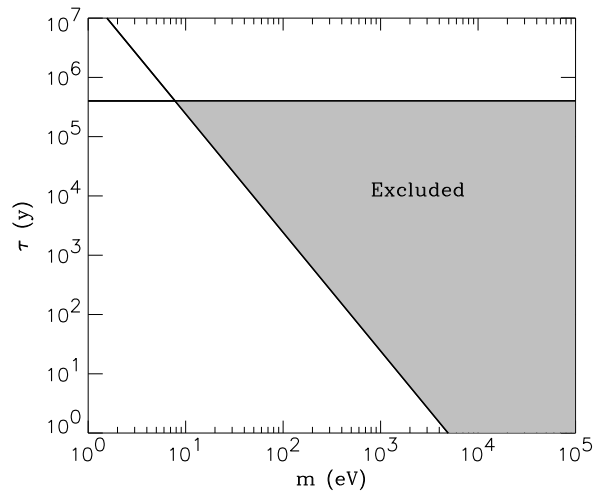}
\vspace{0truecm}
\end{center}
\vspace*{1cm}
\caption{The excluded region for neutrino decays. The horisontal line
indicates the time of recombination. Decays taking place after this
are not constrained by the present calculation. The other full line is
from Eq.~(\ref{eq:alpha}).}
\label{fig2}
\end{figure}

Note that CMBR measurements can also be used to constrain late
neutrino decays, which take place after recombination
\cite{H3,lopez2,lopez3,lopez4}.  However, such decays are detectable
because of the late ISW effect they produce at small $l$.  Since
Boomerang does not detect fluctuations below about $l=50$, we have not
calculated any bound on late decays.

In conclusion, Boomerang has provided us with the first precision CMBR
data. This data can be used to constrain many different types of
exotic physics, not just in the neutrino sector.  The Boomerang data
is only the first indication of what we can expect with the new
generation of CMBR experiments. In the next few years we will have
data from the MAP and Planck satellites \cite{MAP+PLANCK}, which is
expected to be an order of magnitude better than that from Boomerang.

{\it Acknowledgements}--- This letter was supported by the Carlsberg
Foundation. The theoretical power spectra have all been calculated
using the CMBFAST package designed by U.~Seljak and M.~Zaldarriaga
\cite{SZ96}.


\begin{references}
\bibitem{peacock}See for instance J.~A.~Peacock, {\it Cosmological
Physics}, Cambridge University Press (1999).
\bibitem{COBE}G.~F.~Smoot {\it et al.}, Astrophys.\ J.\ {\bf 396}, L1
(1992).
\bibitem{boef}J.~R.~Bond {\it et al.}, Phys.\ Rev.\ Lett.\ {\bf 72},
13 (1994).
\bibitem{jungman1}G.~Jungman {\it et al.}, Phys.\ Rev.\ Lett.\ {\bf
76}, 1007 (1996).
\bibitem{jungman}G.~Jungman {\it et al.}, Phys.\ Rev.\ D {\bf 54},
1332 (1996).
\bibitem{tegmark1}See for instance M.~Tegmark, in: Proc. Enrico Fermi
Summer School, Course CXXXII, Varenna, 1995 (astro-ph/9511148).
\bibitem{bond1}J.~R.~Bond, G.~Efstathiou and M.~Tegmark, Mon.\ Not.\
R.\ Astron.\ Soc.\ {\bf 291}, 33 (1997).
\bibitem{eisenstein}D.~J.~Eisenstein, W.~Hu and M.~Tegmark,
Astrophys.\ J.\ {\bf 518}, 2 (1999).
\bibitem{bernardis}P.~de Bernardis {\it et al.}, Nature {\bf 404}, 955
(2000).
\bibitem{white2000}M.~White, D.~Scott and E.~Pierpaoli,
astro-ph/0004385 (2000).
\bibitem{tegmark2}M.~Tegmark and M.~Zaldarriaga, astro-ph/0004393
(2000).
\bibitem{lisi}E.~Lisi, S.~Sarkar and F.~L.~Villante, Phys.\ Rev.\ D
{\bf 59}, 123520 (1999).
\bibitem{oh}S.~P.~Oh, D.~N.~Spergel and G.~Hinshaw, Astrophys.\ J.\
{\bf 510}, 551 (1999).
\bibitem{H1}S.~Hannestad, Phys.\ Rev.\ D {\bf 61}, 023002 (2000).
\bibitem{SZ96}U.~Seljak and M.~Zaldarriaga, Astrophys.\ J.\ {\bf 469},
437 (1996).
\bibitem{lopez}R.~E.~Lopez {\it et al.}, Phys.\ Rev.\ Lett.\ {\bf 82},
3952 (1999).
\bibitem{kainulai}See for instance R.~Barbieri and A.~Dolgov, Nucl.\
Phys.\ {\bf B349}, 743 (1991); K.~Enqvist, K.~Kainulainen and
M.~Thomson, Nucl.\ Phys.\ {\bf B373}, 498 (1992); R.~Foot and
R.~R.~Volkas, Phys.\ Rev.\ Lett.\ {\bf 75}, 4350 (1995);
D.~P.~Kirilova and M.~V.~Chizhov, Phys.\ Rev.\ D {\bf 58}, 073004 (1998).
\bibitem{H4}S.~Hannestad and G.~Raffelt, Phys.\ Rev.\ D {\bf 59},
043001 (1999).
\bibitem{freese}K.~Freese, E.~W.~Kolb and M.~S.~Turner, Phys.\ Rev.\ D
{\bf 27}, 1689 (1983).
\bibitem{kang}H.-S.~Kang and G.~Steigman, Nucl.\ Phys.\ {\bf B372},
494 (1992).
\bibitem{lesgourgues}J.~Lesgourgues and S.~Pastor, Phys.\ Rev.\ D {\bf
60}, 103521 (1999).
\bibitem{kinney}W.~H.~Kinney and A.~Riotto, Phys.\ Rev.\ Lett.\ {\bf
83}, 3366 (1
999).
\bibitem{LP2000}J.~Lesgourgues and M.~Peloso,
astro-ph/0004412 (2000).
\bibitem{gelmini}G.~Gelmini and A.~Kusenko, Phys.\ Rev.\ Lett.\ 
{\bf 82}, 5202 (1999).
\bibitem{H2}S.~Hannestad, Phys.\ Lett.\ {\bf B431}, 363 (1998).
\bibitem{H3}S.~Hannestad, Phys.\ Rev.\ D {\bf 59}, 125020 (1999).
\bibitem{lopez2}R.~E.~Lopez {\it et al.}, 
Phys.\ Rev.\ Lett.\ {\bf 81}, 3075 (1998).
\bibitem{lopez3}M.~Kaplinghat {\it et al.}, Phys.\ Rev.\ D {\bf 60},
123508 (1999).
\bibitem{lopez4}R.~E.~Lopez, astro-ph/9909414 (1999).
\bibitem{MAP+PLANCK}For more information on these missions, see
the internet pages for MAP (http:///map.gsfc.nasa.gov) and
Planck (http://astro.estec.esa.nl/Planck/).
\end{references}
\end{document}